\documentstyle{article}
 \newcounter{nt}[section]                                  
 \newcounter{nl}[section]                                  
 \date{}                                                   

\begin{document}

\begin{center}
\bf{ON FAST AND SLOW TIMES IN MODELS WITH DIFFUSION}
\end{center}
\vspace{4mm}

\begin{center}
 {\bf M.DE ANGELIS - A.M.MONTE - P.RENNO }
 \footnote{E-mail: modeange@unina.it, amonte@unina.it, renno@unina.it}
\end{center}

\begin {center}
 {\it
  Facolt\`a di Ingegneria, Dip. di Mat. e Appl.,
 via Claudio 21, 80125, Napoli, Italy.}
\end {center}
\vspace{5mm}

{\footnotesize
The linear Kelvin-Voigt operator ${\cal L}_\varepsilon$ is a typical
example of wave operator   ${\cal L}_0$ perturbed  by higher - order viscous
terms as $\varepsilon u_{xxt}$. If   ${\cal P}_\varepsilon$  is a prefixed
boundary - value problem for $ {\cal L}_\varepsilon$,
when $\varepsilon =0$ ${\cal L}
_\varepsilon  $ turns into  ${\cal
L}_0  $
  and ${\cal P}_\varepsilon$ into a problem  ${\cal P}_0$  with the {\em
same} initial - boundary conditions of  ${\cal P}_\varepsilon$.
 Boundary - layers are missing and the related control terms depending
on the {\em fast time} are neglegible. In a {\em small} time -
interval, the wave behavior is a realistic approximation of
$u_{\varepsilon}$ when $ \varepsilon \rightarrow 0$. On the contrary, when
$ t$ is large, diffusion effects should prevail and the behavior of
$u_{\varepsilon}$ for  $ \varepsilon \rightarrow 0$ and  $ t \rightarrow
\infty$ should be analyzed. For this, a suitable functional corrispondence
between  the Green functions  ${\cal G}_\varepsilon$ and   ${\cal G}_0$ of
 ${\cal P}_\varepsilon $ and  ${\cal P}_0$ is achieved and its asymptotic behavior is
rigorously examined.  For this, a suitable functional corrispondence
between  the Green functions  ${\cal G}_\varepsilon$ and   ${\cal G}_0$ of
 ${\cal P}_\varepsilon $ and  ${\cal P}_0$ is derived and its asymptotic behavior is
rigorously examined. As consequence, the interaction between diffusion
effects and pure waves is evaluated by means of the {\em slow time }
$\varepsilon \, t$; the main results show that in time - intervals as
$(\varepsilon , 1/\varepsilon)$ pure waves are propagated nearly
undisturbed, while damped oscillations predominate as from the instant
$ t> 1/\varepsilon$.
}

 {\footnotesize

\vspace{3mm}
  Green's Function, Partial differential equations, Viscoelastic models,
Singular perturbations.

}

  \vspace{6mm}

 \section{\hspace*{-6mm}{\bf .\hspace{2mm}Introduction}}
Consider the class of  dissipative phenomena described by the
following model:

\begin{equation}              \label {11}
{\cal L} _\varepsilon u_\varepsilon \equiv
    (\varepsilon\partial_{xxt} +
c^2 \partial_{xx} - \partial_{tt}) u_\varepsilon  = \ f,\ \  \
\end{equation}

\noindent
where $ \varepsilon, c$ are positive constants and the source term $f $
may be linear or not.
In the linear case, when a prefixed boundary-initial problem  ${\cal P}
_\varepsilon$ is stated, the knowledge of the related  Green function
$G_\varepsilon $ allows to solve explicitly  ${\cal P}
_\varepsilon$. When the function $f$ is not linear, then
$G_\varepsilon $ represents the explicit kernel of the integral equation
to which the problem  ${\cal P}
_\varepsilon$ can be reduced.

\noindent
Two typical examples in the non linear case are:

i) For  $f=a u_t +b \sin u$, the equation (\ref{11}) is the perturbed
Sine-Gordon

\hspace {4mm} equation which models the Josephson effect in
Superconductivity \cite{bp}.

ii) For $f=f(u_x,u_{xt},u_{xx})$,the equation (\ref{11}) is the
Navier-Stokes equation

\hspace{4mm} for a compressible gas with small viscosity
 \cite{mm}.

Moreover, as for the artificial viscosity methods, (\ref{11}) represents a
model of wave equations perturbed by viscous terms with a small
parameter $ \varepsilon$ \cite {kl}, \cite{n}. In this framework, the behaviour of
$u_\varepsilon$ as $\varepsilon \rightarrow 0 $ must be examined and the
interaction of pure waves with the diffusion effects caused by  $\varepsilon u_{xxt}$
 must be estimated.This interaction is meaningful in
the evolution of many dissipative models (viscoelastic liquids or
solids \cite{ta} - \cite{s}, real gases with viscosity \cite{la},
magnetohydrodynamic
fluids \cite{na}.)

For $\varepsilon \equiv 0$, the parabolic equation (\ref{11})
 turns into the wave equation

\begin{equation}              \label {12}
{\cal L} _ 0 u_0  \equiv
    (
c^2 \partial_{xx} - \partial_{tt}) u_0  = \ f,\ \  \
\end{equation}

\noindent
and  ${\cal P}
_\varepsilon$ changes into a problem   ${\cal P}
_0$ for $u_0$, with the {\em same initial-boundary} conditions of  ${\cal P}
_\varepsilon$. So, boundary-layers are missing, but the approximation
of $u_\varepsilon$ by
$u_0$ is rough, because in large time-intervals the diffusion effects are
dominant. As consequence, the asymptotic behaviour of  $u_\varepsilon$
should depend on the  {\em slow time} $\varepsilon \, t$ typical of
the diffusion; on the contrary, boundary layer control terms characterized
by the {\em fast time} $t/\varepsilon$, should be negligible.

Aim of the paper is to derive and analyze an appropriate functional
relation between
 $G_\varepsilon  $ and the wave Green function  $ G_0$ related to
  ${\cal P}
_0$
 (Th. 3.1). By means of this transformation, a rigorous asymptotic
analysis of  $G_\varepsilon$ is achieved (Th. 4.1), and  $G_\varepsilon  $
is approximated by solutions $v(x,t)$ of the second-order
{\em  diffusion-wave} equations

\begin{equation}                         \label{13}
\frac{\varepsilon}{\mit 2} \,  v_{xx}
\  = \ v_t + c v_x,  \ \  \ \ \ \  \\  \  \frac{\varepsilon}{\mit 2} \, v_{xx}
\   = \ v_t - c v_x
  \end{equation}

\noindent
 which correspond to the {\em heat equations}  $v_{yy} \, = \,
 v_{\theta}$, where the time
- variable
 $\theta$  is just the slow - time   $(\theta = \varepsilon \, t) $ and the
space-variables  $y \,= \, x \pm ct$   are related to traveling or
backword waves.

The physical meaning of the above analysis is clarified by the
explicit solution related to the case
 $f=0$ (n.5). Except errors of order $O (\varepsilon /t)$, this relationship
is

\begin{equation}                        \label{14}
u_\varepsilon(x,t) = \frac{c}{\sqrt{2 \pi \varepsilon t}} \
\int_{- \infty} ^{\infty} e^{-\frac{c^2(\tau-t)^2}{2\varepsilon t}}
u_0(x,\tau) \ d\tau,
\end{equation}

\noindent
and it clearly shows the interaction between diffusion and wave
propagation. In the time - interval $(\varepsilon,
\frac{1}{\varepsilon})$, when $\varepsilon \, t < 1$ , pure waves
propagate almost without perturbation; while as from the instant $ t>1/\varepsilon $,
damped oscillations predominate.

Let us remark, generally, asymptotic theories are developed for models
with first and second order operators as documented in \cite{bs} and
\cite{d}. For these operators, in fact, numerous maximum theorems for
the rigorous estimate of the remainder are well known.

As for the third - order model (\ref{11}), the error of the
approximation has been examinated by means of the functional relation
(\ref{31}).

 \section{\hspace*{-6mm}{\bf .\hspace{2mm}Statement of the problem}}
 \setcounter{equation}{0}

\hspace{5.1mm}

If $ u_\varepsilon(x,t)$ is a function defined in the strip\vspace{4mm}

$\ \ \ \ \ \ \ \ \ \ \ \  \ \ \ \ \ \ \   \Omega =\{(x,t) : 0 \leq x
\leq
l, \  \ t>0 \}$,

\vspace{3mm}
\noindent
let ${\cal P}_\varepsilon  $ the initial- boundary value problem related
to the Eq. (\ref{11}) with conditions

  \begin{equation}                                                     \label{21}
  \left \{
   \begin{array}{ll}
   & u_\varepsilon (x,0)=f_0(x), \  \    \partial_t u_\varepsilon
   (x,0)=f_1(x),
   \ \ \  \ x\in [0,l],\vspace{2mm}  \\
    & u_\varepsilon (0,t)=\varphi(t), \  \ u_\varepsilon
    (l,t)=\psi(t), \ \ \ \ \ \ \ \ \ \  \ t>0,
   \end{array}
  \right.
 \end{equation}

\noindent
where  $f_0, f_1, \varphi, \psi$  are arbitrary given data.

Boundary conditions $(\ref{21})_2$ represent only an example of the
analysis we are going to apply; {\em flux-boundary} conditions or {\em mixed
boundary} conditions could be considered too. Further, as $f(x,t),
f_0(x), f_1(x)$ are quite arbitrary, it is not restrictive assuming
$\varphi=0, \ \psi=0$; otherwise, it suffices to put

\vspace{3mm}
\begin{equation}                           \label {22}
\bar{u}_\varepsilon = u_\varepsilon - \frac{x}{l} \psi -
\frac{l-x}{l} \varphi ; \ \ \  \bar{f} = f + \frac{x}{l}
\ddot{\psi} +\frac{l-x}{l} \ddot{\varphi}
\end{equation}

\noindent
and to modify $f_0(x), f_1(x)$ consequently.

Let $\hat{z}(s) $denote the Laplace transform of the function $z(t)$ and
let

\begin{equation}                       \label{23}
\hat{F}(x,s) \ = \ \hat{f}(x,s) - s f_0 (x) - f_1(x).
\end{equation}

\noindent
Further, if $\sigma (\varepsilon) = s ( \varepsilon s + c^2 )^{-1/2}$ ,
let

\begin{equation}             \label {24}
\hat{g}(y,s) = \frac{cosh[(l-y)\sigma]}{2(\varepsilon s +c^2)
\sigma senh (l\sigma)}
\end{equation}

\noindent
and

\begin{equation}                         \label{25}
\hat{G}_\varepsilon(\xi, x, s) = \hat{g} (|x-\xi|, s) -
\hat{g} (|x+\xi|, s).
\end{equation}

\noindent
Then, the Laplace tranform $\hat{u}_\varepsilon$  of the solution of the
problem  ${\cal P}_\varepsilon  $ is given by

\begin{equation}                  \label {26}
\hat{u}_\varepsilon (x,s) = \int_{0}^{l}
[\hat F(\xi,s) +\varepsilon f^{''}_0(\xi)] \hat{G}_\varepsilon
(x,\xi,s) \ d\xi.
\end{equation}

When $\varepsilon =0 $ one has $\sigma(0) = s/c $ and
(\ref{24}),(\ref{25}) turn into

 \begin{equation}             \label {27}
\hat{g}_0(y,s) = \frac{cosh[(l-y)s/c]}{2 \
 c \ s
 \  senh (ls/c)},
\end{equation}

\begin{equation}                         \label{28}
\hat{G}_0(\xi, x, s) = \hat{g}_0 (|x-\xi|, s) -
\hat{g}_0 (|x+\xi|, s).
\end{equation}

\noindent
Problem  ${\cal P}_0  $ is defined by (\ref{12}), (\ref{21}) and the
Laplace transform $\hat{u}_0$ of it's solution is:

\begin{equation}                                      \label{29}
\hat{u}_0 (x,s) = \int_{0}^{l}
\hat F(\xi,s) \\\\\\ \hat{G}_0(\xi,x,s) \ d\xi.
\end{equation}

By comparing (\ref{24}),(\ref{25}) with (\ref{27}), (\ref{28}), the
following relationship between $\hat{G}_\varepsilon$ and
$\hat{G}_0$ is pointed out:

\begin{equation}                                      \label{210}
\hat{G}_\varepsilon(s) = \frac {c^2}{\varepsilon s +c^2} \ \ \hat{G}_0
( \frac{cs}{\sqrt{\varepsilon s+c^2}}
).
\end{equation}

In order to estimate  $G_\varepsilon (t)$ in terms of
$G_0(t)$
 and to achieve the rigorous
 asymptotic behaviour of
$ G_\varepsilon(t)$ for $\varepsilon \rightarrow 0$ ,
 the corrispondence (\ref{210}) must be
analyzed and its inverse Laplace transform must be established.

 \section{\hspace*{-6mm}{\bf .\hspace{2mm} Functional dependence between
$ G_\varepsilon$ and
$ G_0$ }}
 \setcounter{equation}{0}

\hspace{5.1mm}

If  $I_0$ denotes the modified
Bessel function of first kind and order zero, the inverse
Laplace - transform of (\ref{210}) is fully given by the following
theorem:

\vspace {5mm}
{\bf Theorem 3.1}{\em - The relationship between the Green functions}
$ G_\varepsilon(t)$ {\em and}
$ G_0(t)$ {\em is}:
\begin{equation}                                    \label{31}
G_\varepsilon(t)= \frac{c^3
e^{-c^2t/\varepsilon}}{\varepsilon\sqrt{\pi\varepsilon t}} \ \
\int_0^\infty e^{-\frac{c^2v^2}{4\varepsilon t}} \ dv \int_0^v
 G_0(u) I_0 (\frac{2c^2}{\varepsilon} \sqrt{u(v-u)} \ du
\end{equation}

\noindent
{\em and their} ${\cal L}$- {\em transforms satisfy} (\ref{210}) {\em in
the half- plane}
${\it{ Re}}(s) > - c^2/\varepsilon$.

\vspace{6mm}\noindent
{\bf Proof.} For ${\it{ Re}}(\varepsilon s+c^2) > 0$, one has:

\begin{equation}                           \label{32}
{\cal L} ( \frac{e^{-\frac{c^2t}{\varepsilon}}}{\sqrt{\pi \varepsilon
t}}e^{-\frac{c^2v^2}{4\varepsilon t}}) \ \ = \
\frac{e^{\frac{-c}{\varepsilon} \, \sqrt{\varepsilon s +c^2 } \, v }}{\sqrt{\varepsilon
s +c^2}},
\end{equation}

\noindent
therefore the Laplace tranform of (\ref{31}) is :

\begin{equation}                         \label{33}
\hat G_\varepsilon(s)= \frac{c^3
}{\varepsilon\sqrt{\varepsilon s +c^2}}
\int_0^\infty e^{-\frac{c}{\varepsilon}  \sqrt{\varepsilon s +c^2 } \,
v}  dv \int_0^v
 G_0(u) I_0 (\frac{2c^2}{\varepsilon} \sqrt{u(v-u)}  du
\end{equation}
\hspace* {2cm} \[ = \frac{c^3
}{\varepsilon\sqrt{\varepsilon s +c^2}} \ \
\int_0^\infty  G_0(u)  du
 \int_0^\infty e^{-\frac{c}{\varepsilon} (u+v)
\sqrt{\varepsilon s +c^2 }}
 \ I_0 (\frac{2c^2}{\varepsilon} \sqrt{uv}) \ dv .    \]

\vspace{8mm}\noindent
Considering that

\begin{equation}                          \label{34}
 \int_0^\infty e^{-  \frac{cv}{\varepsilon}
\sqrt{\varepsilon s +c^2}}
 \ \ I_0 (\frac{2c^2}{\varepsilon} \sqrt{uv}) \ dv \ =
 \frac{\varepsilon}{c} \ \ \frac{e^{\frac{c^3}{\varepsilon} \frac{u}{
 \sqrt{\varepsilon s +c^2 }}}}{\sqrt{\varepsilon
s +c^2}}
\end{equation}

\noindent
see (\cite{em}), one has

\begin{equation}                            \label{35}
\hat G_\varepsilon(s) = \frac{c^2}{\varepsilon s +c^2} \int_0^\infty
e^{-\frac{cs }{\sqrt{\varepsilon s +c^2}} \ u } \ \ \  G_0 (u) du
\end{equation}

\vspace{8 mm}\noindent
therefore the above  formula coincides with (\ref{210}).
\hbox{} \hfill \rule {1.85mm}{2.82mm}

\vspace {5mm}
Consider now a first application of Theorem 3.1. By means of the
transformation (\ref{31}), it's possible to deduce explicit
expressions for $
G_\varepsilon$  from known formulae of
$ G_0$. So, for instance,
if
$ G_0$ is represented by its Fourier series

\begin{equation}                                     \label{36}
G_0=\frac{2}{c\pi} \ \ \\ \ \sum_{n=1}^{\infty} \ \ \frac{1}{n} \ \
\sin (\frac{\pi}{l}cnt)  \  \sin  (\frac{\pi}{l}nx) \sin
(\frac{\pi}{l} n\xi),
\end{equation}

\noindent
then (\ref{31}) allows to make Fourier series of $G_\varepsilon$
explicit. It suffices to evaluate the transform (\ref{31}) of the
time-component $ G_{0,n}$ of  $ G_{0}$:

\begin{equation}                   \label{37}
 G_{0,n} (t) \ \ = \ \ \sin(\frac{\pi}{l} \ c \ n \ t).
 \end{equation}

Owing to:

\begin{equation}                                     \label{38}
I = \int_{0}^{2v} \sin(ay) I_0(b\sqrt{y(2v-y)})dy =
2 \sin (av) \ \ \frac{ \sin (v \sqrt{a^2-b^2})}{\sqrt{a^2-b^2}},
\end{equation}

\noindent
if one puts:
\begin{equation}
 a=\pi c n/l, \  \ \  b=2c^2/\varepsilon ,  \ \ \  k= 2 c l / (\pi
\varepsilon), \ \ \ \omega_n = \sqrt{1-(n/k)^2},
\end{equation}
 the integral transform (\ref{31}) of  $ G_{0,n}$
gives

\begin{equation}                            \label {39}
G_{\varepsilon,n}(t) \ = \
\  e^{-\frac{\pi^2 n^2 }{2 l^2}  \, \varepsilon \, t }  \
   \frac{\sin (a  \omega_n  t)}
{ \omega_n}.
\end{equation}

\noindent
So, by (\ref{31}),  (\ref{36}),  (\ref{39}), one obtains

\begin{equation}                \label {310}
G_\varepsilon = \frac{2}{\pi c} \sum
_{n=1}^{\infty}e^{-\frac{\pi^2}{2l^2}n^2 \varepsilon t } \sin
(\frac{c \pi n}{l} \omega_n  t
) \frac{1}{n \, \omega_n} \sin  (\frac{\pi}{l}nx) \sin
(\frac{\pi}{l} n\xi).
\end{equation}

\noindent
and this formula represents the Fourier series of  $G_\varepsilon $
previously established in \cite{mda} by
Fourier method.

 \section{\hspace*{-6mm}{\bf .\hspace{2mm} Asymptotic behavior of
 $G_\varepsilon$}}
 \setcounter{equation}{0}

\hspace{5.1mm}

An additional consequence of  (\ref{31}) is the asymptotic
analysis of $G_\varepsilon$ when $\varepsilon$ tends to zero. Referring
to (\ref{31}), let
$\tau=t/\varepsilon$ and let

\begin{equation}                            \label{41}
\Gamma(u,v,\tau)=\frac{(c\sqrt{\tau})^3}{\sqrt{\pi}}
\  G_0(tu) \ I_0(2c^2\tau \sqrt{uv}) e^{
-c^2\tau[1+(u+v)^2/4]}.
\end{equation}

\noindent
Then, the function $G_\varepsilon $ can be given the form

\begin{equation}                            \label{42}
G_\varepsilon =   \  \ \int \! \! \int_{Q} \Gamma(u,v,\tau) du \ \ dv,
\end{equation}

\noindent
where $Q\equiv \{ (u,v)\in [0,\infty]\times[0.\infty] \}.$

So, $G_\varepsilon$ depends on $\varepsilon$ only through the fast time
$\tau$. When $\varepsilon$ is vanishing and $t>\varepsilon$, the parameter
$\tau$ goes to infinity and the asymptotic behaviour of integral
(\ref{42}) can be rigorously obtained by Laplace method \cite{db}

To establish
precise estimates of the remainder terms, let
\begin{equation}                               \label{43}
\chi(\tau)=\chi_0 \ \ \tau^{-1/3},\ \ \ \ \sigma(\tau)=\sigma_0 \ \
\tau^{-1/3},
\end{equation}

\noindent
where  $\chi_0 $ and  $\sigma_0$ are arbitrary real positive constants such
that $\chi$ and $\sigma$ are less then one for all $\tau>1.$ Further, let
\begin{equation}                          \label{44}
Q_0\equiv
\{ (u,v)\in [1-\chi,1+\chi]\times[1-\sigma,1+\sigma] \}  \subset Q.
\end{equation}

\noindent
The following Lemma holds:

\vspace{3mm}\noindent
{\bf Lemma 4.1 -} {\em For all } $\tau>1$, {\em it results}:
\begin{equation}                        \label{45}
|G_\varepsilon - \int \! \! \int_{Q_0} \  \Gamma(u,v,\tau) \,  du\,  dv| < \ \mu
\ e^{-\lambda^2 \tau^{1/3}},
\end{equation}

\noindent
{\em where the constants} $\lambda^2$, $\mu$ {\em depend only on }$ c, \chi_0, $ $\sigma_0$.

\vspace{6mm}\noindent
{\bf Proof.} For all real and positive $z$ one has $I_o(z)<e^z$, so that
by (\ref{41}) one obtains

\begin{equation}                            \label{46}
|\Gamma| \leq  \frac{|G_0|}{\sqrt{\pi}} \ \ (c \sqrt{\tau})^3 \ \ e^{-c^2\tau
\ h(u,v)}
\end{equation}

\noindent
with
\begin{equation}                                \label{47}
h(u,v)=1 \  + \ (1/4)(u+v)^2 \ -2\sqrt{uv}.
\end{equation}

\noindent

\vspace{3mm}
Further, for all $(u,v)\in Q$, one has

\begin{equation}                    \label{48}
h(u,v)=\frac{1}{4}[(u-1)^2+(v-1)^2+(\sqrt u-1)^2 (\sqrt v+1)^2
\end{equation}
\hspace*{2cm} \[
+ (\sqrt v
-1)^2 (\sqrt u +1)^2 ] \ \ \geq \frac{1}{4} [(u-1)^2+ (v-1)^2 ]. \]

\noindent
Therefore, indicating by C a constant independent by  $\tau$, one has:
\begin{equation}                       \label{49}
|\int \! \! \int_{Q-Q_0} \ \Gamma \, du \, dv|\leq  C
 \int \! \! \int_{Q-Q_0}
e^{- \frac{c^2}{4}[(u-1)^2+(v-1)^2]\tau}  \ du \ dv.
\end{equation}

\noindent
Hence, by standard computations, the estimate (\ref{45}) is obtained.
\hbox{} \hfill \rule {1.85mm}{2.82mm}

\vspace{8 mm}
The estimates of
 Lemma 4.1 allow us to define the following
function

\begin{equation}                        \label{410}
H(x,t,\varepsilon ) = \frac{c}{\sqrt{2 \pi \varepsilon t}} \
\int_{- \infty} ^{\infty} e^{-\frac{c^2(\tau-t)^2}{2\varepsilon t}}
G_0(x,\tau) \ d\tau
\end{equation}

\noindent
and to obtain the following result:

\vspace{3mm}\noindent
{\bf Theorem  4.1 -} {\em For all } $(x,t)\in \Omega, $
{\em and} $t>\varepsilon$, {\em one has:}

\begin{equation}                        \label{411}
G_\varepsilon(x,t) = H(x,t,\varepsilon) \ \ (1+ \rho_1) \ + \rho_2,
\end{equation}

\noindent
{\em where the remainder terms} $ \rho _1 , \rho_2 $ {\em are decreasing functions of
the fast time}  $\tau=t/\varepsilon$ {\em such that}

\vspace{3mm}

\begin{equation}                   \label{412}
|\rho _1|  \leq  k_1 \ \tau^{-1}; \ \ \  \ \ \rho_2  \leq \ k_2  \ e^{-
\lambda^2 \tau^{1/3}},
\end{equation}

\noindent
{\em with the constants} $k_1$,  $k_2$, $ \lambda^2$ {\em depending  only on }$ c, \chi_0, $ $\sigma_0$.

\vspace{4mm}
{\bf Proof.} By asymptotic formulae of Bessel functions, for real,
positive and large $z$ one has
(\cite{w}):

\begin{equation}                 \label{413}
I_0(z)=\ \frac {e^z}{\sqrt{2 \pi z}} \ (1+r_0) \ \ \ \ \\ \  (|r_0| < c_0
\ \ z^{-1})
\end{equation}

\noindent
and so the integral of $\Gamma$  on $Q_0 $ (see (\ref{41}), (\ref{42}))
can be given the form

\begin{equation}                   \label{414}
\int \! \!  \int_{Q_0} \\ \Gamma du dv = \frac{c \tau}{2 \pi} \
\int_{1-\chi}^{1+\chi}
G_0(tu) \, du  \int_{1-\sigma}^{1+\sigma} \ \frac{1+r_1}{(uv)^{1/4}} \ \
 e^{- c^2 \tau \ h(u,v)} \ dv ,
\end{equation}

\noindent
where $h(u,v)$ is defined in (\ref{47}) and the remainder term $r_1$ is
such that $|r_1| \, \leq c_1 \,( \tau \sqrt{uv})^{-1}$, according to
(\ref{413}).

The  function h(u,v) has a point of absolute minimum in
$(u_0,v_0)=(1,1)$, where it vanishes.
Therefore, it results: $h\, =\,h_0 \, +\,h_*$ with

\begin{equation}            \label{415}
h_0=\frac{1}{2}[(u-1)^2+(v-1)^2]; \ \ \ \ \   |h_*| \leq c_*
[|u-1|+|v-1|]^3.
\end{equation}

\noindent

According to the Laplace method,  \cite{db}, for large $\tau$ , the dominant
contribution is due to a neighbourhood of $(u_0,v_0)$ , so that it's
enough to expand $\Gamma$ near to this point. Further, if $u$ and $v$ are
substituted by

\begin{equation}                 \label{416}
u  - 1 \ = \ \frac{1}{\sqrt{\tau}} \ u_1, \ \ \ v  - 1 \ = \
\frac{1}{\sqrt{\tau}} \ v_1,
\end{equation}

\noindent
the terms of third order like $h_*$ are vanishing with order of $\tau
^{-3/2}$ ; besides, the limits of integration become $( -\infty, \
+\infty)$ as:   $\chi \sqrt{\tau}=\chi_0 \ \tau^{1/6}$ and $\sigma
\sqrt{\tau}=\sigma_0 \ \tau^{1/6}$.

Then, for $\tau \rightarrow \infty$, one has:

\begin{equation}                  \label {417}
\int \! \! \int_{Q_0} \Gamma dudv = \frac{c^2 }{2\pi} \int
_{-\infty}^{\infty} e^{-\frac{c^2}{2} u^2}  G_0(t+\sqrt{\varepsilon t} \  u) du
\int
_{-\infty}^{\infty} e^{-\frac{c^2}{2}  v^2} dv+
 O(\frac{1}{\tau}).
\end{equation}

\noindent
The above formula, together with the results of Lemma 4.1
(see(\ref{45}),(\ref{410})), implies (\ref{41}) and the estimate
$(\ref{412})_2$ for $\rho_2$. Finally, by means of routine calculations,
the order of the error  $\rho_1$ can be rigorously specified according to
$(\ref{412})_1$.
\hbox{} \hfill \rule {1.85mm}{2.82mm}

 \section{\hspace*{-6mm}{\bf .\hspace{2mm} Diffusion and waves}}
 \setcounter{equation}{0}

\hspace{5.1mm}

To remark the physical meaning of the results, let consider the simple
case: $ f=0$, $f_0=0, f_1 \neq 0.$ By (\ref{26}) and (\ref{29}) one
has:

\begin{equation}                          \label{51}
u_\varepsilon = - \int_0^l f_1 (\xi)G_\varepsilon (\xi,x,t) d\xi ; \
\ \ \ u_0 = - \int_0^l f_1(\xi) G_0 (\xi,x,t) d\xi
\end{equation}

\noindent
Consequently, except errors of order $O(\varepsilon /t) $, Theorem 4.1 allows to
obtain (see (\ref{410}),(\ref{411})) the following estimate:

\begin{equation}                        \label{52}
u_\varepsilon(x,t) = \frac{c}{\sqrt{2 \pi \varepsilon t}} \
\int_{- \infty} ^{\infty} e^{-\frac{c^2(\tau-t)^2}{2\varepsilon t}}
u_0(x,\tau) \ d\tau.
\end{equation}

\noindent
The above explicit corrispondence between $u_0$ and $u_\varepsilon$ clearly
shows the interaction between diffusion and wave propagation.

For instance, when $ f_1= \frac{c\pi}{l} \ \sin (\frac{\pi x}{l})$, one has
the {\em pure wave}

\begin{equation}                 \label{53}
u_0(x,t)= \sin (\frac{\pi x}{l}) \ \ \sin (\frac{\pi }{l} \ c t ),
\end{equation}

\noindent
while $u_\varepsilon$, owing to (\ref{52}), is given by

\begin{equation}                 \label{54}
u_\varepsilon(x,t)=  e^{-\frac{\pi^2  }{2 l^2}  \, \varepsilon \, t }
\\
\ u_0(x,t).
\end{equation}

\vspace{3mm}

\noindent
This means that the dissipation caused by $\varepsilon u_{xxt}$ is
significant by slow
times  $\varepsilon \, t$ only; the terms depending on the fast time
$t/\varepsilon$ are fully negligible for all $t >\varepsilon$.

\noindent
As consequence, in the time interval $(\varepsilon,
\frac{1}{\varepsilon})$ where $\varepsilon t <1$, the pure wave (\ref{53})
is propagated nearly undisturbed; when $t>\frac{1}{\varepsilon}$, damped
oscillations predominate.

These aspects are generally valid, whatever the initial data may be,
and for appropriate $f$.  In fact,
if one transforms by (\ref{410}) the time-components
$ G_{0,n} (t) = \sin(\frac{n\pi}{l} \, c \, t)$  of $G_0$,
one has \cite{em}:

\begin{equation}                        \label{55}
 \frac{c}{\sqrt{2 \pi \varepsilon t}} \
\int_{- \infty} ^{\infty} e^{-\frac{c^2}{2} \, \frac{(\tau-t)^2}{\varepsilon t}}
\sin (\frac{\pi}{l} \, c \, n \, \tau) \ d\tau =  e^{-\frac{\pi ^2
n^2}{2l^2}
\,
\varepsilon \,  t} \ \sin (\frac{\pi}{l} \, c \, n \, t).
\end{equation}

\noindent
Therefore, the Fourier series of the function $H$ which approximates
$G_\varepsilon$  is given by

\begin{equation}                         \label{56}
H =
\frac{2}{c\pi} \ \ \\ \ \sum_{n=1}^{\infty} \ \ \frac{1}{n} \ \
 e^{-\frac{1}{2}(\frac{\pi n}{l})^2 \, \varepsilon \, t } \   \sin (\frac{\pi}{l}cnt)  \  \sin  (\frac{\pi}{l}nx) \sin
(\frac{\pi}{l} n\xi).
\end{equation}

This formula, together with (\ref{26}), (\ref{411}),   makes explicit the
asymptotic behavior of the solution $u_\varepsilon $ of the problem $\cal
P_\varepsilon$ related to arbitrary data $f_0,f_1$ and $f$. As (\ref{56})
shows, this behavior is like (\ref{54}), with damped waves which
become vanishing as from the instants  $t> 1/\varepsilon$.

It must be remarked that the function $H $ can be given the form:

\noindent
$H=H^--H^+$, with

\begin{equation}                         \label{57}
H^{\pm} =
\frac{2}{c\pi} \sum_{n=1}^{\infty}  \frac{1}{n} \
 e^{- \frac{1}{2} (\frac{\pi n}{l})^2 \ \varepsilon  t }  \   \sin (\frac{\pi n}{l}\xi)
 \cos [\frac{\pi n}{l}(x \pm ct)].
\end{equation}

\noindent
Then, the basic variables which are typical of the diffusion and wave
propagation are identified by the following

\vspace{4mm}
\noindent
{\bf Property 5.1} - {\em The function}  $H^-$ {\em (or} $H^+${\em ) is solution of
the diffusion-wave equation}

\begin{equation}                         \label{58}
\frac{\varepsilon}{2} v_{xx}
\ = \ v_t + c v_x  \ \ \ \ \ (or : \frac{\varepsilon}{2} v_{xx}
\ = \ v_t - c v_x).
  \end{equation}

\noindent
{\em Moreover, if one puts}:   $y=x \pm ct, \ \ \ \theta =
\frac{\varepsilon}{2} \,t$, {\em Eq.} (\ref{58}) {\em turns
into the heat equation} $v_{yy} =v_{\theta}$, {\em with the time
-variable}
$\theta$ {\em given just by the slow time and the
space-variable} $y$ {\em related to traveling (or retrograde) waves.}

\vspace {5mm}
Finally, we observe that $H^{\pm}_x $ is given by
Jacobi  Theta function    $\theta_3(y,\theta)$;
in fact, for  $y=x + ct$ or  $y=x - ct$, one has:

\begin{equation}                        \label{59}
H^{\pm}_x = \frac{1}{4cl} [ \theta_3 (\frac{y-\xi}{2l},
\frac{\theta}{2l^2})-
\theta_3 (\frac{y+\xi}{2l
}, \frac{\theta}{2l^2})],
\end{equation}

\noindent
according to well-known formulae related to the heat equation.

\vspace {10mm}

\noindent

\vspace{5mm}
\bf{REFERENCES}

\end {document}